\documentclass{icrc2009}
\usepackage{graphicx}   % for including figures
\usepackage[caption=false]{caption}    % for captions
\usepackage[font=footnotesize]{subfig} % subfig.sty for a double column floating figure using two subfigures
\usepackage{url}

\newcommand{\shorttitle}[1]%
{\markboth{Proceedings of the 31\MakeLowercase{$^{st}$} ICRC, {\L}\'{o}d\'{z} 2009}{#1} }
\newcommand{\etal}{\MakeLowercase{\textit{et al. }}} % "et al."

\begin{document}
\title{An Alignment System for Imaging Atmospheric\\Cherenkov Telescopes}

\author{\IEEEauthorblockN{A. McCann, D. Hanna and M. McCutcheon} \\ \IEEEauthorblockA{Department of Physics, McGill
    University, H3A 2T8, Montreal, QC, Canada
    (mccann@physics.mcgill.ca)}}

% please write the preseter's name and short title (3-4 words maximum)
%    which will appear at the header of the even pages.
\shorttitle{McCann \etal Alignment System for IACT\MakeLowercase{s}}
\maketitle

\begin{abstract}
The reflector used by an imaging atmospheric Cherenkov telescope
(IACT) consists of a tessellated array of mirrors mounted on a large
frame. This arrangement allows for a very large reflecting surface
with sufficient optical quality for the implementation of the IACT
technique at a moderate price. The main challenge presented by such a
reflector is maintaining the optical quality, which depends on the
individual alignment of several hundred mirror facets. We describe a
method of measuring and correcting the alignment of the mirror facets
of the reflectors used by the VERITAS telescopes. This method employs
a CCD camera, placed at the focal point of the reflector, which
acquires a series of images of the reflector while the telescope
performs a raster scan about a star. Well-aligned facets appear bright
when the telescope points directly at the star while misaligned facets
appear bright when the angle between the telescope pointing direction
and the star is twice the misalignment angle of the mirror. Data from
these scans can therefore be used to produce a set of corrections
which can be applied to the facets. In this contribution we report on
initial experience with an alignment system based on this principle.
\end{abstract}

 
\section{Introduction}
Ground-based gamma-ray astronomy is a rapidly developing field in
observational science which is being driven by the current generation
of imaging Cherenkov telescopes operating around the world
\cite{Holder,Hinton,Baixeras}. This generation of telescopes has built
on its predecessors with improvements in several aspects of detector
design, including the use of larger reflectors and larger cameras with
a greater number of PMT elements. For the performance afforded by
these improvements to be fully realised, the quality of the
telescopes' optical alignment must be at a high level. The VERITAS
array, located at the base of Mount Hopkins in southern Arizona,
consists of four 12 meter diameter Davies-Cotton reflectors. Each
reflector is composed of $\sim$345 hexagonal mirror facets, all of
which must be properly aligned for the telescope optics to be
optimal.\\ The exposure of the mirrors to the dust of the Arizona
desert requires their periodic re-coating to maintain optimal
reflectivity \cite{Roache}. The drawback of this re-coating procedure
is that facets are regularly removed and replaced and this leads to
the degradation of the telescope alignment. The telescopes must
therefore be regularly realigned. This provided the stimulus to find a
quick and easily implemented method of characterising and correcting
the alignment of the telescopes without resorting to a costly active
mirror control system. The method reported here, called the {\em
  raster method}, is based on the SCCAN technique described by
Arqueros \etal \cite{Arqueros}.

\section{The Raster Method}
\begin{figure}[!t]
\centering
\subfloat[Star on axis]{\includegraphics[width=2.8in]{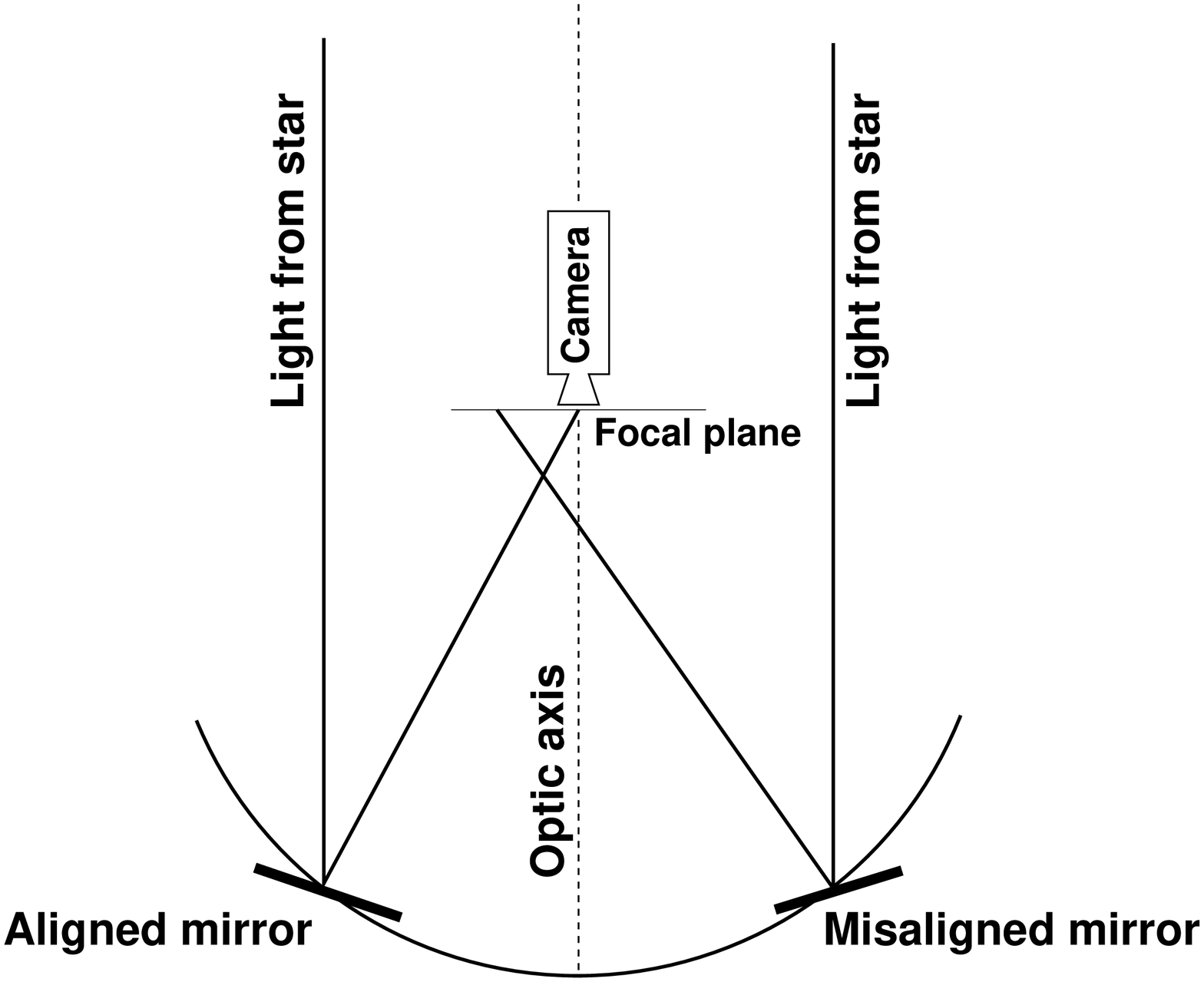}}

\subfloat[Star off axis]{\includegraphics[width=2.8in]{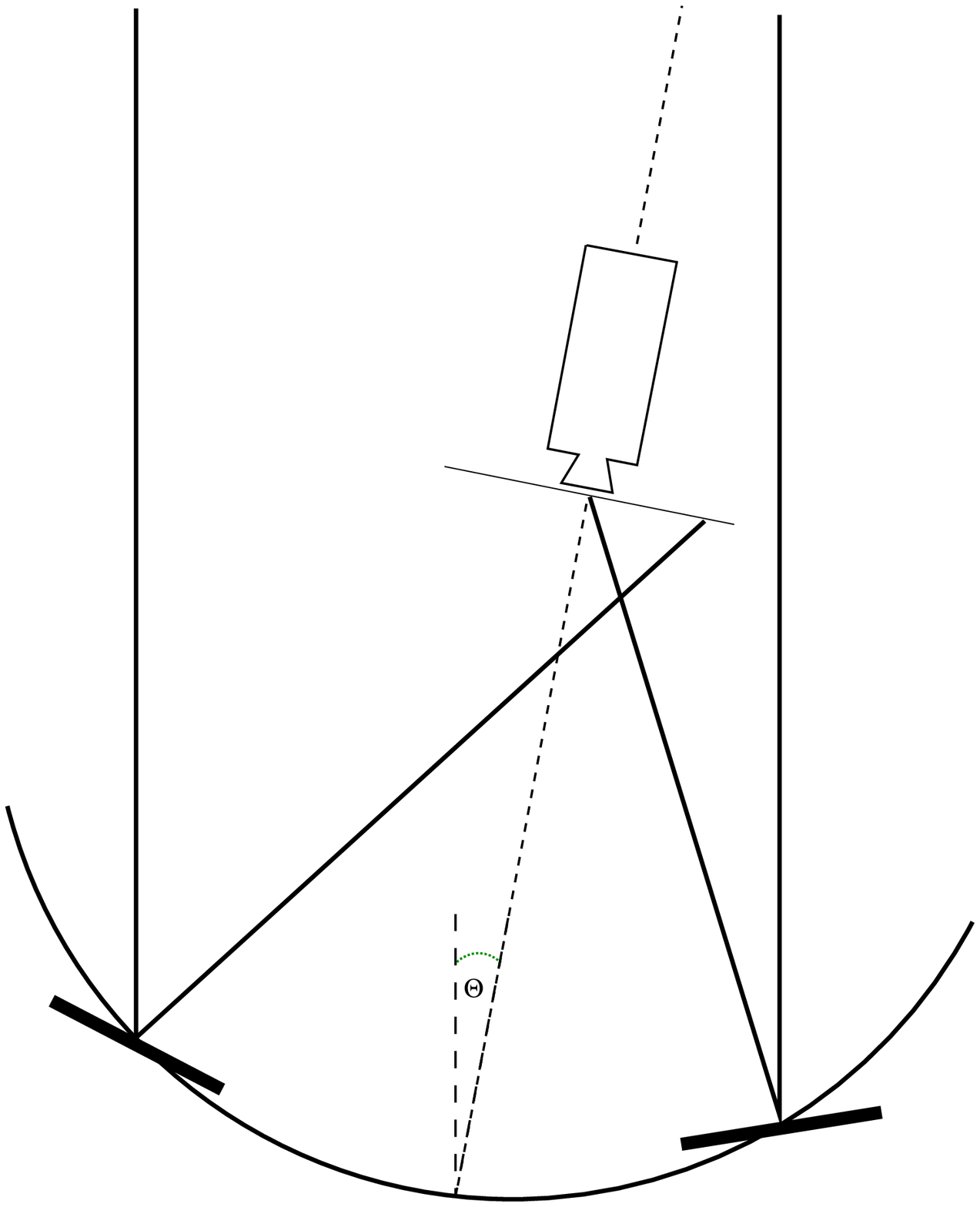}}
\caption{An illustration of the raster scan method. In figure (a) the
  well-aligned facet will appear bright in the CCD image while the
  misaligned facet will be dark. In figure (b) misaligned facet will
  appear bright when the angle between the star and the telescope
  pointing direction is twice the misalignment angle of the facet.}
\label{fig:cartoon}
\end{figure}

\begin{figure}[!t]
  \centering
  \includegraphics[width=2.8in]{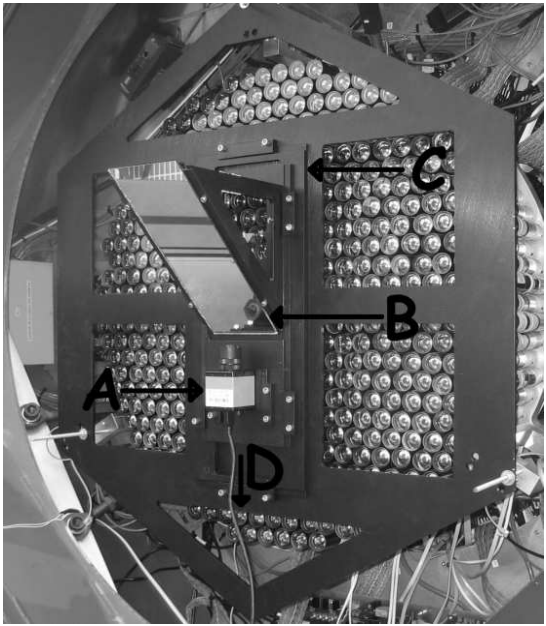}
  \caption{A photograph of the alignment apparatus mounted on one of
    the VERITAS cameras. Arrow {\em A} indicates the CCD camera; {\em
      B}, the 45 degree mirror; {\em C}, the x-y positional stage and
    {\em D} the wire connecting to the data acquisition laptop (out of
    shot).}
  \label{fig:app}
\end{figure}
\begin{figure}[!t]
\centering
\subfloat[]{\includegraphics[width=2.7in]{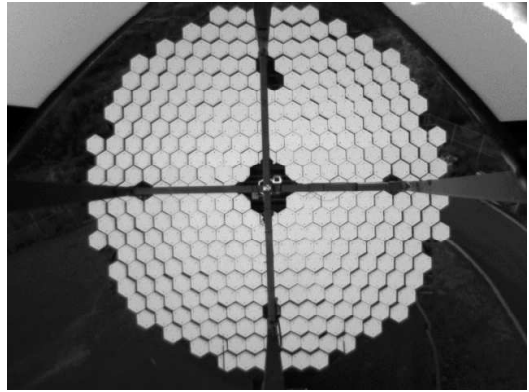}}

\subfloat[]{\includegraphics[width=2.7in]{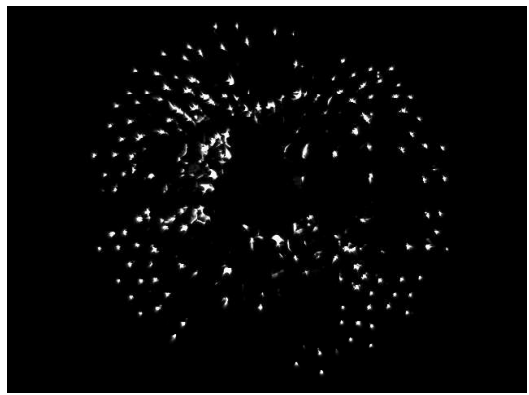}}
\caption{Images, take by the CCD camera, of one of the VERITAS
reflectors during the day (a) and at night, while pointing at a
star (b). The bright spots in image (b) are caused by starlight
reflecting off well-aligned facets while the dark regions indicate
poorly-aligned facets.}
\label{fig:dish}
\end{figure}

The {\em raster method} employs a CCD camera which is placed at the
focal point of the telescope, facing the reflector. The telescope is
made to track a star at a typical observation elevation ($\sim$70
degrees). The telescope is then made to perform a raster scan around
the position of the star. This has the effect of moving the image of
the star across the CCD camera. At each point in the raster scan the
CCD camera records how much starlight each facet is contributing to
the image. Well-aligned facets will contribute their maximum amount of
light when the telescope is pointing directly at the star. Mis-aligned
facets will contribute their maximum amount of light when the angle
between the star and the telescope pointing direction is twice the
misalignment angle of the facet (see
Figure~\ref{fig:cartoon}). Performing a scan of this type allows for
the determination of the misalignment angle, and therefore the
required adjustments, for every facet of the telescope.\\ A photograph
of the apparatus used to implement this method is shown in Figure
~\ref{fig:app}. The main components of the apparatus are:
 \begin{itemize}
   \item a mounting plate
   \item an x-y positional stage
   \item a 45-degree mirror
   \item a CCD camera with wide-angle lens
   \item a laptop computer
\end{itemize}
The mounting plate and x-y positional stage are made from anodised
aluminium. They provide a sturdy base for the CCD camera, allowing it
to be mounted on the optic axis of the telescope, directly in front of
the VERITAS PMT camera. The 45 degree mirror is positioned such that
the virtual image of the CCD camera is placed at the focal plane of
the reflector. The CCD camera is an {\em Imaging Source} DMK 21BF04
camera with a 1/4 inch 640x480 monochrome CCD and a firewire
interface. The lens is an {\em Imaging Source} H0514-MP lens with a
5mm focal length. Image acquisition software runs on a {\em Dell}
Inspiron laptop which is connected to the CCD camera and, via an
ethernet cable, to the telescope tracking computer. Images of the
telescope reflector taken by the CCD camera are displayed in
Figure~\ref{fig:dish}.

\section{Initial test}
\begin{figure}[!t]
  \centering
  \includegraphics[width=2.7in]{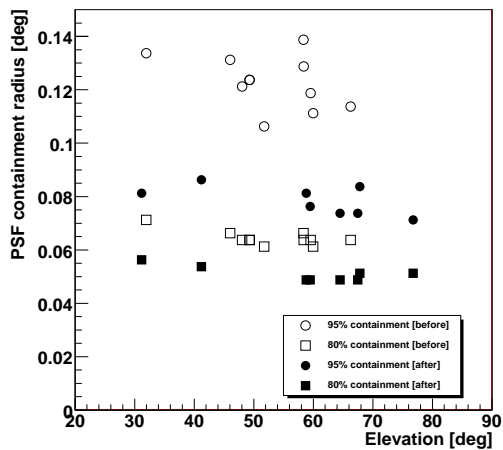}
  \caption{A plot of the Point Spread Function of the VERITAS
    telescope 2 before and after alignment with the raster
    system.}
  \label{fig:psf}
 \end{figure}

This alignment system was first used on one of the VERITAS telescopes
in April 2009. VERITAS Telescope 2, which had undergone the most facet
replacements over the last year and thus had the worst Point Spread
Function (PSF) among the telescopes in the array, was chosen for the
first test. A star of magnitude 2, which had a transit elevation of 68
degrees, was tracked over a two hour period. During this time the
average elevation was 66.5 degrees. The telescope was made to raster
scan over a square grid of 21x21 points, with a step size of 0.025
degrees. From these data a set of adjustments was calculated. Of the
345 facets mounted on the dish, 150 were deemed to require
adjustment. The mirrors of the VERITAS telescope are mounted on the
optical support structure via a triangular bracket with a threaded rod
at each vertex, supporting a mounting gimbal and adjustable nut. Any
misalignment of a particular facet can be corrected by turning two of
%the adjustable nuts \enlargethispage{-28.0mm} on the mirror mount
the adjustable nuts on the mirror mount
assembly. Adjustments were made to 120 of the mirrors with a wrench
which had a circular index wheel attached to it, so that adjustments
of 1/2,1/4,1/8 and 1/16 of a turn were easily applied. The 30
remaining mirrors were located at the very top of the dish and could
not be accessed at the time of adjustment due to high winds. PSF
measurements were made before and after the adjustments were
implemented and are plotted in Figure~\ref{fig:psf}. The 80\%
containment radius of the PSF at 66 degrees elevation decreased from
0.064 degrees to 0.049 degrees after the alignment, a 22\%
improvement. Further improvement is expected with the adjustment of
the 30 remaining mirrors and an iteration of measurement and
adjustment based on a raster scan undertaken with greater angular
resolution.

\section{Discussion}
The initial test of the {\em raster method} has proved successful. The
raster scan was completed in two hours and the data categorising the
mirror alignment have been shown to be both useful and accurate. The
procedure has proved to be much easier to implement than the previous
alignment method \cite{Toner} and indeed more accurate. This initial
test has not, however, investigated the limit of the precision which
is possible. Further tests are scheduled in which we will perform
raster scans with a smaller grid spacing. With the current system the
density of sampling across the raster grid, and therefore the accuracy
to which we can measure the alignment of the telescope, must be
balanced against the amount of time it takes to perform the scan. The
characterisation of the telescope alignment must be made while the
telescope is tracking over a narrow elevation band
($\sim$5 degrees) due to the deformation of the optical support
structure caused by gravitational slump. In the next scheduled
alignment test we will perform a raster with high granularity over a
raster grid comparable to the size of the, now reduced, PSF image. To
enable us to implement the mirror adjustments calculated from a high
resolution raster scan we are developing a ``geared'' wrench with a
ratio of four turns to one. This will allow us to reliably make
adjustment as small as $\sim$0.009 degrees (corresponding to 1/32nd of
a turn of the adjustable nut) to each mirror. Ray-tracing simulations
with the commercial package, {\em FRED}, and an independent ray-tracing
analysis suggest that a PSF with an 80\% containment radius of 0.042
degrees should be attainable within the telescope specifications.

\section{Conclusion}
An alignment system based on the SCCAN method suggested by
\cite{Arqueros} has been built, and tested on one of the VERITAS
telescopes. The system has performed very well in this initial
test. Further investigations are scheduled.

\section{Acknowledgements}
VERITAS is supported by grants from the US Department of Energy, the
US National Science Foundation, and the Smithsonian Institution, by
NSERC in Canada, by Science Foundation Ireland, and by STFC in the
UK. We acknowledge the excellent work of the technical support staff
at the FLWO and the collaborating institutions in the construction and
operation of the instrument. The authors gratefully acknowledge the
help of Ryan Irvin, Ken Gibbs, Stephen Fegan, Jack Musser, Jeremy
Perkins, Stephanie Wissel, Victor Acciari, Gary Gillanders and Mark
Lang.

\end{document}